\documentclass[english]{article}
\usepackage[T1]{fontenc}
\usepackage[latin9]{inputenc}
\usepackage{color}
\usepackage{textcomp}
\usepackage{amstext}
\usepackage{amssymb}
\usepackage{stackrel}
\usepackage{amsmath}
\usepackage{graphicx}
\usepackage{esint}
\usepackage{cite}
\usepackage{epstopdf}
\usepackage{epsfig}
\usepackage{subfigure,array}

\makeatletter

\providecommand{\keywords}[1]
{
	\small	
	\textbf{\textit{Keywords---}} #1
}

\usepackage[a4paper]{geometry}
\makeatother

\usepackage{babel}
\begin{document}

\title{Nonclassicality in Two-Mode New Generalized Binomial State}
\author{Kathakali Mandal$^{\dagger,\mathsection}$,Anjali Jatwani$^{\mathsection}$and Amit Verma$^{\mathsection,}$\\$^{\mathsection}$Jaypee Institute of Information Technology,Sector-128,Noida,UP-201304,India\\$^{\dagger}$ Center for Molecular Spectroscopy and Dynamics, Institute for Basic Science(IBS),
\\Korea University, Seoul 02841,Korea}
\maketitle
\begin{abstract}
The study of nonclassical properties of two-mode quantum states is particularly useful in quantum information theory because of the possibilities of obtaining entanglement and other two-mode quantum correlations in these states. Here we have investigated the possibilities of the existence of nonclassicality in a two-mode New generalized binomial state (TMNGBS). Specifically two-mode antibunching, Quadrature squeezing, sum \& difference squeezing, and various entanglement criteria e.g Shchukin-Vogel entanglement criterion, the uncertainty relation of SU(1,1) Algebra and EPR entanglement criterion are explored in two mode particular example of quantum state named as New generalized binomial state. Earlier we studied nonclassicality in single-mode NGBS, here we are extending our study toward the two-mode version of a quantum state. Here we provide the general expressions of moments for a two-mode quantum state (Fock basis) and explore the quantification in a particular example NGBS. It is found that antibunching, squeezing, and SV entanglement are possible with different limits of depending parameters but the entanglement criteria (EPR, SU (1,1) algebra and Cauchy - Schwarz inequality based)for NGBS are not possible. This study opens up the possibility of exploring the two-mode nonclassicality in other states too. 
\end{abstract}
\keywords{Nonclassical states, two-mode New generalized binomial state, antibunching, entanglement, SU(1,1) algebra.}

\section{Introduction}
In quantum state engineering\cite{miranowicz2004dissipation} and quantum computation,
nonclassical properties of quantum states are very useful and an important field of study \cite{chen2019quantifying,chen2018simulating,alam2019bose, barnett2018statistics,verma2010generalized}.
This is so because nonclassical states have no classical analogue
and can thus be useful in realizing impossible tasks
in the classical world and can only establish quantum supremacy \cite{harrow2017quantum}.
Examples of nonclassical properties are squeezing, antibunching, and entanglement. In summary, nonclassical features of quantum states
are very important and the same has been studied for various families of quantum states \cite{verma2019study,mandal2019generalized,malpani2020impact,mandal2019higher,deepak2024comparative}. In most of the earlier works on quantum states, nonclassicality in single mode state has been explored majorly but nonclassicality in two modes is relatively new \cite{alam2017lower,baghshahi2015generation,han2019time,vintskevich2020entanglement,ren2023nonclassical}. Two modes or multi photon nonclassicality finds applications in quantum computing and quantum communications\cite{ren2019nonclassical,dat2021non,hertz2022decoherence}. In particular, quantum cryptography is solely dependent on entangled photons. Thus, there is a natural interest in studying entanglement properties of the states of many photons. Recent experiments have also shown that entanglement of up to ten photons can be observed in the laboratory \cite{vintskevich2020entanglement,truong2021detecting}. \textcolor{black}{Here it would be apt to note that a few interesting experiments involving qudits have been performed in the recent past. The number of such successful experiments are small compared to experiments involving qubits, but the experiments are related to various aspects of physics including but not restricted to superconducting phase qudit, quantum football, quantum state tomography of large nuclear spin, and time-domain grating with a periodically driven qutrit \cite{han2019time}}. In \cite{franco2007generating}, Franco et al. proposed an unconditional Scheme to generate the N-photon
quantum superposition of two orthogonal binomial states of electromagnetic field for the two-photon case in a single mode high-Q cavity by using resonant atom-cavity interaction which shows the possibility of experimental generation of qudit states. \textcolor{black}{Similarly, many works on nonclassicality in different systems are reported in the literature. For example, unified derivation for multi-mode nonclassicality \cite{miranowicz2010testing}, sudden vanishing and reappearance of nonclassical effects in a system, increasing relative nonclassicality quantified by standard entanglement potentials \cite{miranowicz2015statistical} and more recently quantifying the nonclassicality of pure dephasing are reported in literature \cite{chen2018simulating,chen2019quantifying}}. On the other hand, qudit ($d$-level states) states are defined as a finite superposition of Fock states in $d$-dimensional Hilbert space \cite{miranowicz2004dissipation,mandal2020higher,alam2018higher}. A particular subclass of qudit states is the set of finite-dimensional intermediate states \cite{verma2010generalized,verma2008higher,mandal2020higher}. One such intermediate state named as a new generalized binomial state (NGBS) was
introduced by Fan et.al \cite{fan1999new} and we have recently reported many results for nonclassicality in single mode NGBS \cite{mandal2019generalized}. The effect of single photon addition/subtraction and multiphoton addition/subtraction are also investigated by us in recent reports \cite{mandal2019higher,mandal2020higher,wang2016nonclassicality,swain2022two,ren2022nonclassical}. Fan et.al \cite{fan1999new} also show that the factor $q$ plays an essential role in the displaying highly nonclassical behaviors of NGBS and it is the generalization of the Binomial states(BS) as for $q=0$, NGBS converts into BS which can further be reduced into most classical coherent state and most nonclassical Fock states and vacuum states in different limits of depending parameters as shown in Fig. 1.
 \begin{figure}
\begin{center}
\includegraphics[scale=0.5]{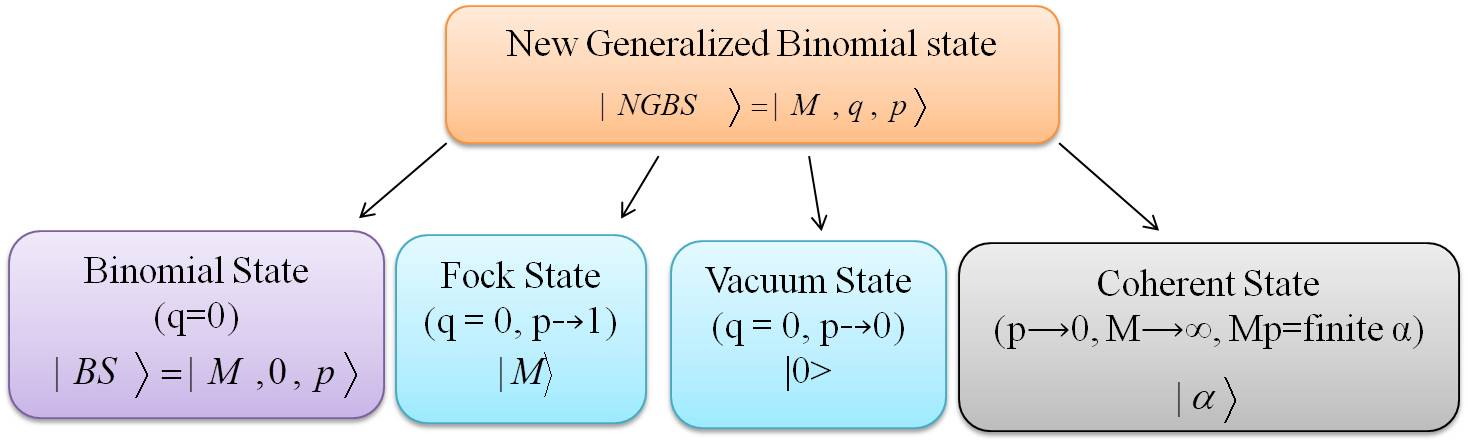}
\caption{(color online) A schematic diagram for the generalized nature of NGBS which converts in to BS, Fock state, vacuum state, and coherent state with different limits of depending parameters. Here p is probability, q is cavity factor, and M is the total number of photons available in NGBS.}
\end{center}
\end{figure}

After Fan et.al work on single-mode NGBS, Wang et.al \cite{wang2019time} introduced a two-mode version of NGBS (TMNGBS) as Generalized two-mode binomial states \cite{dong2009entanglement}, which may be generated by using experimental techniques. Wang et.al investigated the entanglement of the relative entropy of TMNGBS and then explored the teleportation application by using the mean fidelity of the scheme. Recently we have reported higher-order nonclassicality and the single photon source criteria with the help of probability distribution in single mode (NGBS) \cite{mandal2019generalized,mandal2020higher,Jatwani2023}. Here in this paper, motivated by the results of Wang et. al, we want to study higher-order nonclassicality and various Entanglement properties in TMNGBS as a specific example of a two-mode quantum state. For the same we investigate the possibilities of the existence of two-mode antibunching \cite{an2002multimode}, quadrature squeezing and sum and difference squeezing and three kinds of entanglement criteria named as Shchukin and Vogel criterion \cite{shchukin2005nonclassicality}, Uncertainty relation of SU(1,1) algebra criterion \cite{nha2006entanglement} and finally EPR entanglement criterion given by Mancini et.al \cite{mancini2002entangling}. This paper is organized in 4 sections. In section 2, we define the two-mode quantum state in general and show the analytic form of the moment. Section 3 presents various types of entanglement criteria and two-mode nonclassical criteria in the form of antibunching and squeezing and finally section 4 is dedicated to the conclusion.

\section{Two-mode finite dimensional Quantum state: Moment calculation}

 A two-mode quantum state can be written as
\begin{equation}
\psi=\sum_{n=0}^{M}C_{n}|n\rangle |M-n\rangle\label{eq:psi},
\end{equation}
where $C_{n}$ is the probability amplitude, $|n\rangle$ and $|M-n\rangle$ represents Fock state having $n$ and $M-n$ photon respectively. Here we consider $|n\rangle$ as one mode with creation and annihilation operators $a_{1}^{\dagger}$ and $a_1$ respectively. Similarly  $|M-n\rangle$ as another mode with creation and annihilation operators $a_{2}^{\dagger}$ and $a_2$ respectively throughout this paper. Further repeated operation of creation and annihilation operators provide general expressions of normally ordered moment quadrature for both the modes which are as follows,
moment for mode $|n\rangle$ 
\begin{equation}
\langle a_{1}^{\dagger k}a_{1}^l \rangle_{n, l>k}=\sum_{n=l}^{M}C_{n}C_{n-l+k}\left[\frac{n!(n-l+k)!}{(n-l)!^2}\right]^{\frac{1}{2}}\label{eq:akal1},
\end{equation}
\begin{equation}
\langle a_{1}^{\dagger k}a_{1}^l \rangle_{n, l<k}=\sum_{n=l}^{M-(k-l)}C_{n}C_{n-l+k}\left[\frac{n!(n-l+k)!}{(n-l)!^2}\right]^{\frac{1}{2}}\label{eq:akal2},
\end{equation}
and moment for mode $|M-n\rangle$
\begin{equation}
\langle a_{2}^{\dagger k}a_{2}^l \rangle_{M-n, l>k}=\sum_{n=0}^{M-l}C_{n}C_{n+l-k}\left[\frac{(M-n)!(M-n-l+k)!}{(M-n-l)!^2}\right]^{\frac{1}{2}}\label{eq:akal3},
\end{equation}
\begin{equation}
\langle a_{2}^{\dagger k}a_{2}^l \rangle_{M-n, l<k}=\sum_{n=k-l}^{M-l}C_{n}C_{n+l-k}\left[\frac{(M-n)!(M-n-l+k)!}{(M-n-l)!^2}\right]^{\frac{1}{2}}\label{eq:akal4}.
\end{equation}
Here Eq.(\ref{eq:akal1}-\ref{eq:akal4}) describe the general form of a moment for any two-mode quantum state. Further, with the help of these equations and a particular probability amplitude $C_{n}$ of any quantum state, we can explore the possibilities of entanglement and other nonclassical properties. For this purpose, we consider $C_{n}$ for TMNGBS which is defined as 
\begin{equation}
C_{n}=\left[\frac{p}{(1+Mq)}\frac{M!}{n!(M-n)!}\left(\frac{p+nq}{1+Mq}\right)^{n-1}\left(1-\frac{p+nq}{1+Mq}\right)^{M-n}\right]^{\frac{1}{2}}\label{eq:ngbs1}
\end{equation}
All the criteria of nonclassicality and Entanglement can be analytically studied by using the above equations. So in the next section, we first study the different criteria of nonclassicality and then various Entanglement criteria for the analytical study of TMNGBS.

\section{Nonclassicality in two mode New generalised binomial state \label{sec:entanglement criteria:}}
In this section, we first study various two modes nonclassicality criteria (specifically two-mode antibunching, Quadrature squeezing, sum squeezing) in TMNGBS. Then in later subsections, we explored various Entanglement criteria in TMNGBS. Entanglement
was always been a fundamental property. For the two-mode system, a state is to be considered entangled if it cannot be decomposed into two subsystems distinctly. In recent years, it has been
reported that various types of entangled states may be generated by the two common continuous-variable type states, such as coherent states and squeezed states. Here we wish to consider three criteria of entanglement, which are as follows:
\subsection{Two-mode antibunching}
Antibunching phenomenon in quantum information is essential for making a single-photon source, which is highly useful in quantum cryptography. Earlier we have studied single mode higher order antibunching in many quantum states \cite{verma2008higher} including single mode NGBS too \cite{mandal2019generalized}. Here we are interested to explore HOA in TMNGBS as the signature of nonclassicality. Originally, a criterion of higher order antibunching was given by Lee \cite{lee1990many} which is given below for two arbitrary modes x and y $(x \neq y = {a_1,a_2})$, 
\begin{equation}
\langle n_{x}^{(l+1)}n_{y}^{(m-1)}+n_{y}^{(l+1)}n_{x}^{(m-1)}\rangle < \langle n_{x}^{(l)}n_{y}^{(m)} +n_{y}^{(l)}n_{x}^{(m)}\rangle\label{eq:anti1}
\end{equation}
where $l\geq m \geq 1$. For two modes $x$ and $y$, any quantum state is said to be antibunched of the order of $l, m$ if $D_{l,m}(2) < 0$, where
\begin{equation}
D_{l,m}(2)= \langle n_{x}^{(l+1)}n_{y}^{(m-1)}+n_{y}^{(l+1)}n_{x}^{(m-1)}\rangle / \langle n_{x}^{(l)}n_{y}^{(m)} + n_{y}^{(l)}n_{x}^{(m)}\rangle -1.\label{eq:anti2}.
\end{equation}
Here, number operator in respective modes are defined as $n_{x}= a_{1}^{\dagger}a_{1}$ and $n_{y} = a_{2}^{\dagger}a_{2}$. Finally expression, in terms of creation and annihilation operators of respective modes, is reduced to 
\begin{equation}
D_{l,m}(2)= \langle a_{1}^{\dagger l+1}a_{1}^{l+1}a_{2}^{\dagger m-1}a_{2}^{m-1}+a_{2}^{\dagger l+1}a_{2}^{l+1}a_{1}^{\dagger m-1}a_{1}^{m-1}\rangle/\langle a_{1}^{\dagger l}a_{1}^{l}a_{2}^{\dagger m}a_{2}^{m}+a_{2}^{\dagger l}a_{2}^{l}a_{1}^{\dagger m}a_{1}^{m}\rangle-1\label{eq:anti3}
\end{equation}
\begin{figure}
\centering
\subfigure[]{\includegraphics[scale=0.75]{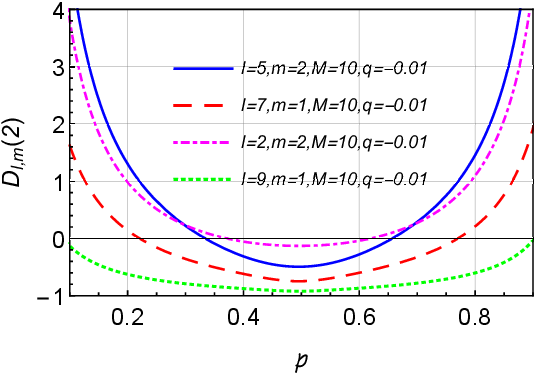}}\quad\subfigure[]{\includegraphics[scale=0.75]{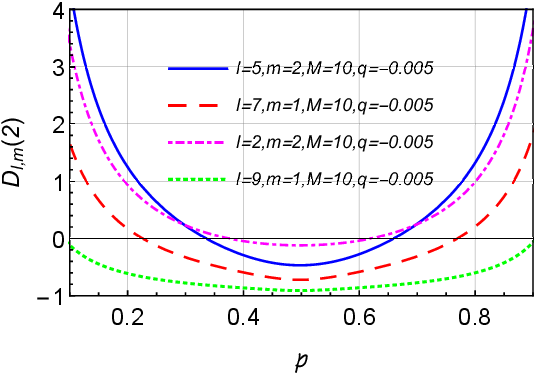}}\quad\subfigure[]{\includegraphics[scale=0.75]{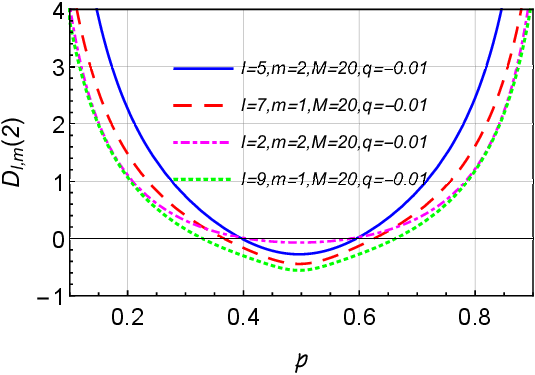}}\quad\subfigure[]{\includegraphics[scale=0.75]{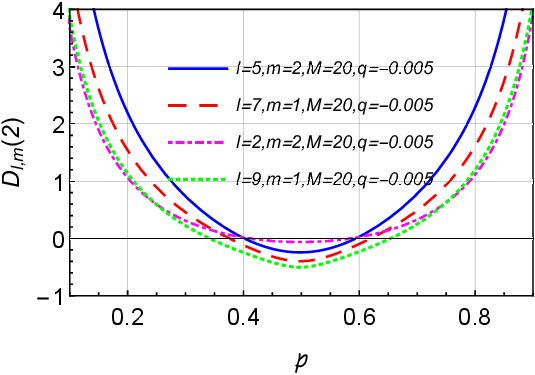}}\\
\caption{\label{fig:HOA}(Color online) 2(a) and 2(b) show two mode higher order antibunching for NGBS corresponding to M=10 and q=-0.01, q=-0.005 respectively, for various combinations of l and m. 2(c) and 2(d) show two mode higher order antibunching for NGBS corresponding to M=20 and q=-0.01, q=-0.005 respectively, for various combinations of l and m. HOA is prominent around the mean value of p (approx range 0.4 to 0.6 in 2(c) and 2(d). The range of HOA with respect to p increases with least values of M and high difference in l,m as shown in 2(a) and 2(b).}
\end{figure}
In Fig. \ref{fig:HOA}, variation of HOA for two-mode NGBS is shown with probability p and different values of M, q, and various combinations of orders $(l,m)$. It is observed that nonclassicality depth increases with the difference between $l$ and $m$. For example, in 2(a)-2(d), depth of HOA is less for $l=2, m=2$ and more for $l=9, m=1$. It is also noticed that M should be the least, q must be more negative and depth maximizes around p=0.5. HOA is absent for positive values of q so our earlier claim of negative q satisfies in the two-mode study too. 
\subsection{Quadrature Squeezing}
\begin{figure}
\centering
\subfigure[]{\includegraphics[scale=0.75]{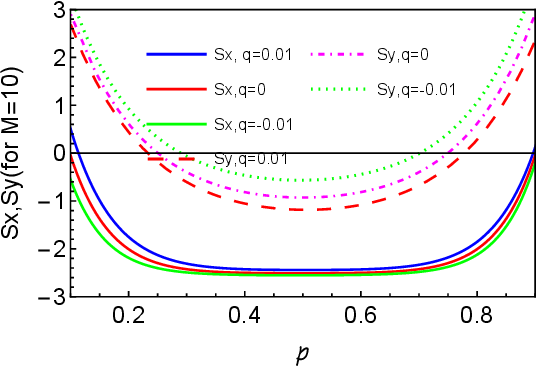}}\quad\subfigure[]{\includegraphics[scale=0.75]{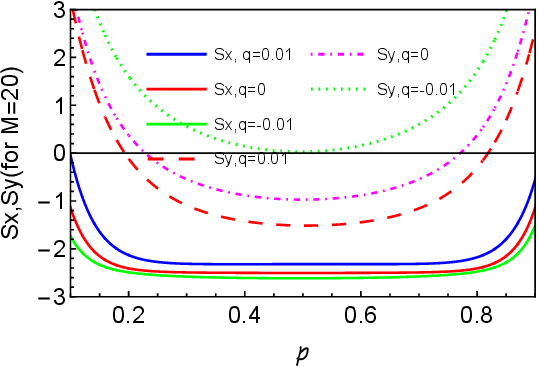}}\\
\caption{\label{fig:sxy}(Color online)  (a) and (b) are showing two-mode quadrature squeezing for NGBS corresponding to M=10, M=20 respectively for various values of q = 0.01, q = 0, q = -0.01 (corresponding to the curve marked by solid blue, solid red, and solid green respectively for $S_{x}$ and dash red line, dash pink line, and dotted green respectively for $S_{y}$). $S_{x}$ is more prominent than $S_{y}$, and the negativity of QS increases with increasing M. Further positive value of q has more negative $S_{x}$ and $S_{y}$.}
\end{figure}

Here we consider another criterion of nonclassicality for two-mode as Quadrature Squeezing (QS), which can be measured by homodyne detection technique. When the state is superimposed on a strong coherent beam of local oscillator\cite{ren2019entanglement}, The two-mode QS can be observed and quadrature operators are defined as
\begin{equation}
X_{1,2}=\frac{1}{2\sqrt{2}}(a_{1}^{\dagger}+a_{1}+a_{2}^{\dagger}+a_{2})\label{eq:sx}
\end{equation}
\begin{equation}
Y_{1,2}=\frac{i}{2\sqrt{2}}(a_{1}^{\dagger}-a_{1}+a_{2}^{\dagger}-a_{2})\label{eq:sy}.
\end{equation}
The degree of squeezing is best parameterized by
\begin{equation}
\langle (\delta X_{1,2})^2\rangle < \frac{1}{4}    \hspace{5pt} or  \hspace{5pt}  \langle (\delta Y_{1,2})^2\rangle < \frac{1}{4}\label{eq:sxy1}.
\end{equation}
Where $\langle (\delta X_{1,2})^2\rangle = \langle X_{1,2}^2\rangle - \langle X_{1,2}\rangle^2$
Substituting (\ref{eq:sx}) and (\ref{eq:sy}) in Eq. (\ref{eq:sxy1}) the squeezing factors associated with $X_{1,2}$  and $Y_{1,2}$ can be expressed as 
\begin{equation}
S_{x}=Re\langle a_{1}^{2}+a_{2}^{2}+2(a_{1}a_{2}^{\dagger}+a_{1}a_{2})\rangle + \langle a_{1}a_{1}^{\dagger}+a_{2}a_{2}^{\dagger}\rangle -2Re^{2}\langle a_{1} + a_{2}\rangle-2 <0, \label{eq:s2x}
\end{equation}
\begin{equation}
S_{y}= -Re\langle a_{1}^{2}+a_{2}^{2} - 2(a_{1}a_{2}^{\dagger}- a_{1}a_{2})\rangle+\langle a_{1}a_{1}^{\dagger}+a_{2}a_{2}^{\dagger}\rangle - 2Im^{2}\langle a_{1} + a_{2}\rangle-2<0, \label{eq:s2y}
\end{equation}
In Fig.\ref{fig:sxy}(a) and \ref{fig:sxy}(b), Eq.(\ref{eq:s2x}) and Eq.(\ref{eq:s2y}) are plotted as two-mode QS for NGBS with values of M = 10 and M = 20 respectively, with different values of p and q. It is observed that $S_{x}$ is more prominent than $S_{y}$, and the negativity of QS increases with increasing M. Further, $S_{x}$ is more negative for negative values of q in comparison to positive values, and this behavior is reversed in the case of $S_{y}$ for positive and negative values of q. 
\subsection{Sum squeezing}
\begin{figure}
\centering
\subfigure[]{\includegraphics[scale=0.75]{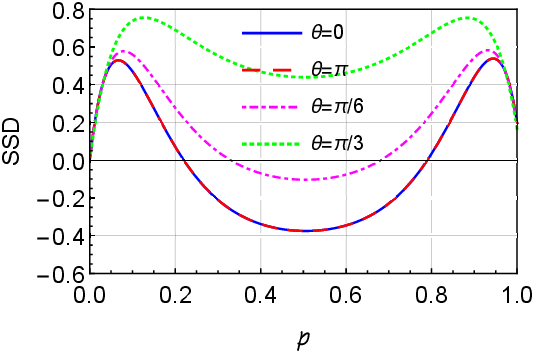}}\quad\subfigure[]{\includegraphics[scale=0.75]{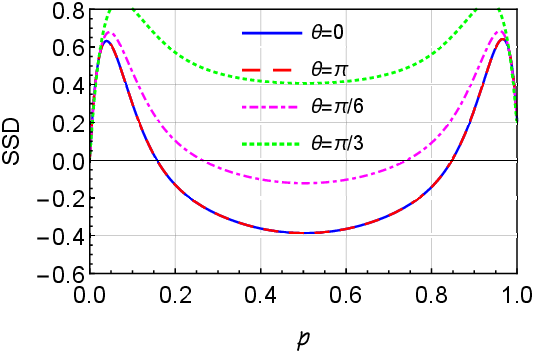}}\\
\caption{\label{fig:ssd1}(Color online) (a) and (b) are showing two-mode sum squeezing for NGBS corresponding to M = 10 and M = 20, respectively, for various values of $\theta = 0 $, $\theta = \pi $, $\theta = \frac{\pi}{6} $, $\theta = \frac{\pi}{3} $. For  $\theta = 0 $ and $\theta = \pi $ the value of SSD is same which is more negative. SSD is more prominent for $\theta = 0 $ and $\theta = \pi$.}
\end{figure}

Sum and difference squeezing are two multi-mode versions of higher-order squeezing. These two types of squeezing become normal single-mode squeezing by using the appropriate nonlinear optical process. For example, sum squeezing can be converted into normal squeezing via sum-frequency generation, and difference squeezing realized the same by difference frequency. Thus we can detect these forms of squeezing in the experiment via this property. The sum squeezing is defined by a two-mode operator in the form as
\begin{equation}
V_{e}=\frac{1}{2}(e^{i\theta}a_{1}^{\dagger}a_{2}^{\dagger}+e^{-i\theta}a_{1}a_{2}) \label{ssd1}
\end{equation}
Where $\theta$ is an angle, $\theta = 0$ and $\theta = \pi/2$ corresponding to the real and imaginary parts of the operator $V_{e}$, respectively.
A two-mode state is said to be sum squeezed if  
\begin{equation}
SSD=\frac{4\langle (\Delta V_{e}^{2}\rangle - \langle a_{1}a_{1}^{\dagger} + a_{2}a_{2}^{\dagger}-1\rangle )}{\langle a_{1}a_{1}^{\dagger}+a_{2}a_{2}^{\dagger}-1\rangle}<0, \label{ssd2}
\end{equation}
where $\langle (\Delta V_{e})^2\rangle = \langle V_{e}^2\rangle-\langle V_{e}\rangle^2$
Substituting Eq. (\ref{ssd1}) into Eq. (\ref{ssd2}) we have the degree of the sum squeezing ($SSD$) in the form of antinormally ordered operators as following
\begin{equation}
SSD= \frac{2\langle a_{1}a_{1}^{\dagger}a_{2}a_{2}^{\dagger} \rangle +2 Re(e^{-2i\theta}\langle a_{1}^{2}a_{2}^{2} \rangle)-4Re^{2}(e^{-\theta}\langle a_{1}a_{2}\rangle)}{\langle a_{1}a_{1}^{\dagger}+a_{2}a_{2}^{\dagger}-1 \rangle}-2<0
\end{equation}
\\
In Fig.\ref{fig:ssd1}, the two mode squeezing for NGBS is shown for the values of M = 10 and M = 20 with the different values of  $\theta = 0 $, $\theta = \pi $, $\theta = \frac{\pi}{6} $, $\theta = \frac{\pi}{3} $. The graph indicates that when $\theta$ equals 0 or $\pi$, the quantity SSD exhibits a greater negative value. The more negative value of the quantity SSD indicates a stronger degree of squeezing present in the system. 
\subsection{Shchukin-Vogel Entanglement}
\begin{figure}
\centering
\subfigure[]{\includegraphics[scale=0.75]{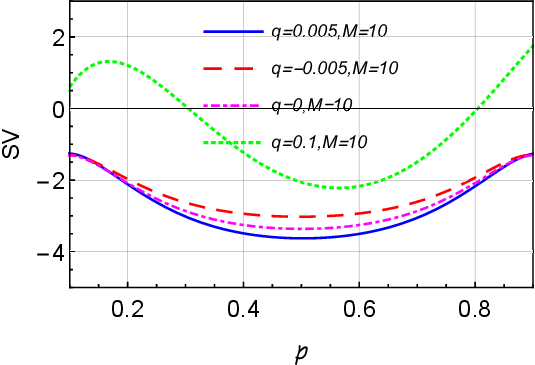}}\quad \subfigure[]{\includegraphics[scale=0.75]{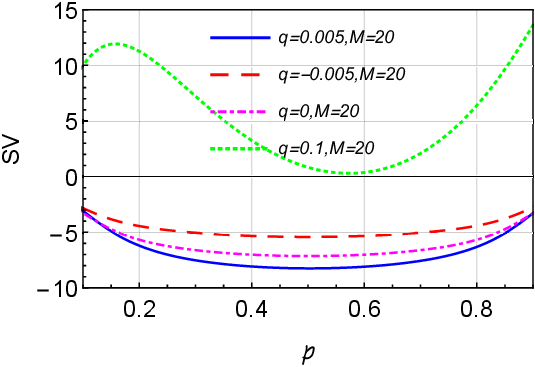}}\\
\caption{\label{fig:SV}(Color online) (a) and (b) are showing SV criterion for TMNGBS as a function of probability p with different values of M = 10, q = 0.01, -0.01, 0.0, 0.1, and M = 20, q = -0.005, 0.0, -0.005, 0.1, respectively.}
\end{figure}

Shchukin et al.\cite{shchukin2005inseparability} presented a sufficient and necessary condition for the inseparability of the two-mode quantum states. This sufficient Shchukin-Vogel (SV) condition for entanglement was originally given in anti-normal ordering form as given below 
\begin{equation}
SV=\langle a_{1}a_{1}^{\dagger}-\frac{3}{2}\rangle \langle a_{2}a_{2}^{\dagger}-\frac{3}{2}\rangle -\langle a_{1}^{\dagger}a_{2}^{\dagger}\rangle \langle a_{1}a_{2}\rangle <0\label{eq:SV1}.
\end{equation} 
After normal ordering, we can get
\begin{equation}
SV=\langle a_{1}^{\dagger}a_{1}-\frac{1}{2}\rangle \langle a_{2}^{\dagger}a_{2}-\frac{1}{2}\rangle -\langle a_{1}^{\dagger}a_{2}^{\dagger}\rangle \langle a_{1}a_{2}\rangle <0
\label{eq:SV2}
\end{equation} 
By using Eq. (\ref{eq:SV2}), we study entanglement in TMNGBS. Fig.\ref{fig:SV} (a) and (b) show the SV criterion in TMNGBS is plotted for M = 10 and M = 20, respectively, with different values of the depending parameters q and p. It is observed that for least values (positive) of q, the depth of negative region is more. 

\subsection{Einstein-Podolsky-Rosen (EPR) criteria }
Mancini et al. \cite{ren2019entanglement,wang2018quantifying} provided suitable entanglement criteria and demonstrated the robustness of the achieved entanglement against thermal noise. To describe the entanglement between the two modes, the condition for the entangled states or inseparable states is given as
\begin{equation}
EPR = I_1I_2 - 1 <0\label{eq:EPR1},
\end{equation}
where 
\begin{equation}
I_1 = Re \langle a_1 ^{2}+ a_2 ^{2} +2 a_1a_2+2a_1a_2^{\dagger}\rangle + \langle a_1a_1^{\dagger}+ a_2^{\dagger}a_2\rangle + 2Re^{2} \langle a_1+a_2 \rangle - 1\label{eq:i1},
\end{equation} and
\begin{equation}
I_2 = Re \langle -a_1 ^{2}- a_2 ^{2} +2 a_1a_2-2a_1a_2^{\dagger}\rangle + \langle a_1a_1^{\dagger}+ a_2a_2^{\dagger}\rangle + 2Im^{2} \langle a_1-a_2\rangle - 1\label{eq:i2},
\end{equation}
Eq.\ref{eq:EPR1} defines the criterion for determining the presence of entanglement, which we have employed in our analysis of the two-mode state NGBS. However, it fails to show the entanglement behavior in two-mode NGBS.

\subsection{The criteria in terms of SU(1,1) algebra  }
This criterion is based on
the uncertainty of relation of SU(1, 1) algebra.The angular momentum operators are defined as\cite{ren2019entanglement,wang2018quantifying}
\begin{equation*}
J_x = \frac{1}{2} \left( a_1^{\dagger}a_2 + a_1a_2^{\dagger}\right)\label{eq:jx}
\end{equation*} 
\begin{equation*}
J_y = \frac{1}{2} \left( a_1^{\dagger}a_2 - a_1a_2^{\dagger}\right)\label{eq:jy}
\end{equation*}
\begin{equation}
J_z = \frac{1}{2} \left( a_1^{\dagger}a_1 - a_2^{\dagger}a_2\right)\label{eq:jz}  
\end{equation}
which have the canonical commutation relations $\left[J_x , J_y\right]= iJ_x$ and cyclic permutations. The uncertainty relation in the SU$\left(1,1\right)$ algebra is 
\begin{equation}
\Delta J_x \Delta J_y \geq \frac{1}{2}|\langle J_x\rangle|.
\end{equation}
 \\
 Using this property of partial transpose and Eq.
(23), the following inequality is also obtained

\begin{equation}
\begin{aligned}
S_U &= \left[2 \langle a_1a_1^{\dagger}a_2a_2^{\dagger}\rangle - \langle a_1a_1^{\dagger}\rangle - \langle a_2a_2^{\dagger}\rangle \right. \\
&\quad + 2\mathrm{Re} \left( a_1^{2}a_2^{\dagger 2}\right) - 4\mathrm{Re}^{2} \langle a_1^{\dagger}a_2\rangle  \Big] \\
&\quad \times \left[2 \langle a_1a_1^{\dagger}a_2a_2^{\dagger}\rangle - \langle a_1a_1^{\dagger}\rangle - \langle a_2a_2^{\dagger}\rangle \right. \\
&\quad \left. - 2\mathrm{Re} \left( a_1^{2}a_2^{\dagger 2}\right) - 4\mathrm{Im}^{2} \langle a_1^{\dagger}a_2\rangle  \right] \\
&\quad - |\langle a_1a_1^{\dagger}- a_2a_2^{\dagger}\rangle|^{2}\geq0
\end{aligned}
\end{equation}
\\
A violation of Eq. (25) would imply that the state is
entangled. It can be noted that we have applied this criterion to assess the entanglement of the two-mode NGBS in this paper however, we did not get the expected results. We have also explored Cauchy-Schwarz inequality and covariance measures of entanglement in two-mode NGBS. A summary of our observation is mentioned in Table.1.

\section{Conclusion\label{sec:CONCLUSION}}
Here we have investigated the possibilities of the existence of higher-order nonclassicality in TMNGBS of radiation. Specifically, two-mode antibunching and entanglement properties with the help of several criteria are explored, for example, the Shchukin-Vogel entanglement criterion, the SU(1,1) algebra-based entanglement criterion, and EPR criterion of entanglement given by Mancini et al. The two-mode quantum state studied here is easy to generate experimentally using optical elements. Motivated by these facts and potential applications of two-mode quantum correlations in quantum information processing, we have analytically studied higher-order nonclassical properties of the two-mode new generalized binomial state, and have established that this state can be used for various quantum information processing tasks.
\begin{table}
\begin{center}
\begin{tabular}{|m{8cm}|m{6cm}|}
\hline
Nonclassicality Criteria & Present\\
\hline
Higher order two-mode antibunching & Yes\\
\hline
Quadrature squeezing & Yes\\
\hline
Sum and difference squeezing & Yes \\
\hline
Entanglement criteria &\\
\hline
Shchukin-Vogel criteria&Yes\\
\hline
EPR criteria & No \\
\hline
The criteria in terms of SU(1,1) algebra & No \\
\hline
Cauchy-Schwarz inequality & No\\
\hline
Covariance measures of entanglement & No\\
\hline
\end{tabular}
\end{center}
\caption{Table.1: Table shows all nonclassicality criteria studied in this paper irrespective of the presence or nonpresence of nonclassicality in TMNGBS.}
\end{table}

\section*{Acknowledgment} AV acknowledges the support from the QuEST scheme of the Interdisciplinary Cyber-Physical Systems (ICPS) program of the Department of Science and Technology (DST), India (Grant No.: DST/ICPS/QuST/Theme-1/2019/14 (Q80)). The authors also acknowledge Prof. Anirban Pathak for their constant support and encouragement.

\bibliographystyle{unsrt}
\bibliography{QSartical4}

\end{document}